\documentclass[fleqn,usenatbib]{mnras}

\usepackage{newtxtext,newtxmath}
\usepackage[T1]{fontenc}
\usepackage{ae,aecompl}

\usepackage{graphicx}	% Including figure files
\usepackage{amsmath}	% Advanced maths commands
\usepackage{amssymb}	% Extra maths symbols
\usepackage{ulem}
\usepackage{siunitx}

\title[A 40.6-Minute Eclipsing Double White Dwarf System]{ZTF J1901+5309: A 40.6-Minute Orbital Period Eclipsing Double White Dwarf System}

\author[M. W. Coughlin et al.]{   Michael W. Coughlin,$^{1,2}$\thanks{E-mail: cough052@umn.edu}
Kevin Burdge,$^{2}$
E. Sterl Phinney,$^{2}$
Jan van Roestel,$^{2}$
\newauthor
Eric C. Bellm,$^{3}$
Richard G. Dekany,$^{4}$
Alexandre Delacroix,$^{4}$
Dmitry A. Duev,$^{2}$
\newauthor
Michael Feeney$^{4}$
Matthew J. Graham,$^{2}$
S. R. Kulkarni,$^{2}$
Thomas Kupfer,$^{5}$
\newauthor
Russ R. Laher,$^{6}$
Frank J. Masci,$^{6}$
Thomas A. Prince,$^{2}$
Reed Riddle,$^{4}$
\newauthor
Philippe Rosnet,$^{7}$
Roger Smith,$^{4}$
Eugene Serabyn,$^{8}$
Richard Walters$^{4}$
\\
$^{1}$School of Physics and Astronomy, University of Minnesota, Minneapolis, Minnesota 55455, USA\\
$^{2}$Division of Physics, Math, and Astronomy, California Institute of Technology, Pasadena, CA 91125, USA\\
$^{3}$DIRAC Institute, Department of Astronomy, University of Washington, 3910 15th Avenue NE, Seattle, WA 98195, USA\\
$^{4}$Caltech Optical Observatories, California Institute of Technology, Pasadena, CA 91125, USA\\
${5}$Kavli Institute for Theoretical Physics, University of California, Santa Barbara, CA 93106, USA\\
$^{6}$IPAC, California Institute of Technology, 1200 E. California Blvd, Pasadena, CA 91125, USA\\
$^{7}$Universit{\'e} Clermont Auvergne, CNRS/IN2P3, LPC, Clermont-Ferrand, France\\
$^{8}$Jet Propulsion Laboratory, California Institute of Technology, Pasadena, CA 91109, USA
}

\date{Accepted XXX. Received YYY; in original form ZZZ}

\pubyear{2019}

\begin{document}
\label{firstpage}
\pagerange{\pageref{firstpage}--\pageref{lastpage}}
\maketitle

\begin{abstract}
The Zwicky Transient Facility has begun to discover binary systems with orbital periods that are less than 1\,hr.
Combined with dedicated follow-up systems, which allow for high-cadence photometry of these sources, systematic confirmation and characterization of these sources are now possible.
Here, we report the discovery of ZTF J190125.42+530929.5, a 40.6\,min orbital period, eclipsing double white-dwarf binary.
Both photometric and spectroscopic modeling confirm its nature, yielding an estimated inclination of $i = 86.2^{+0.6}_{-0.2}\,\rm degrees$ and primary and secondary effective temperatures of $\textrm{T}_\textrm{eff} = 28,000^{+500}_{-500}\,K$ and $\textrm{T}_\textrm{eff} = 17,600^{+400}_{-400}\,K$ respectively.
This system adds to a growing list of sources for future gravitational-wave detectors and contributes to the demographic analysis of double degenerates.
\end{abstract}


\section{Introduction}

\begin{figure*}
 \includegraphics[width=3.2in]{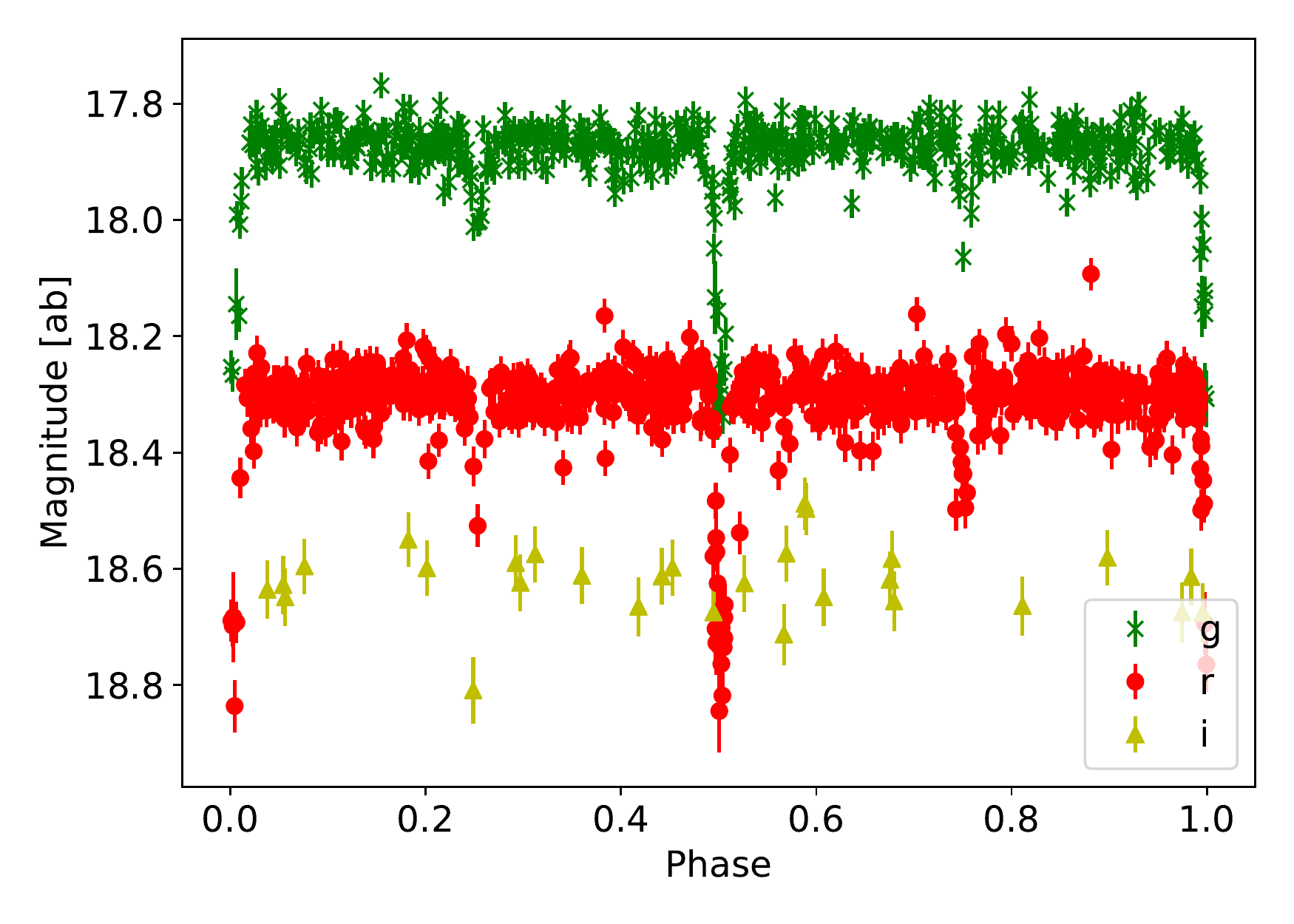}
  \includegraphics[width=3.2in]{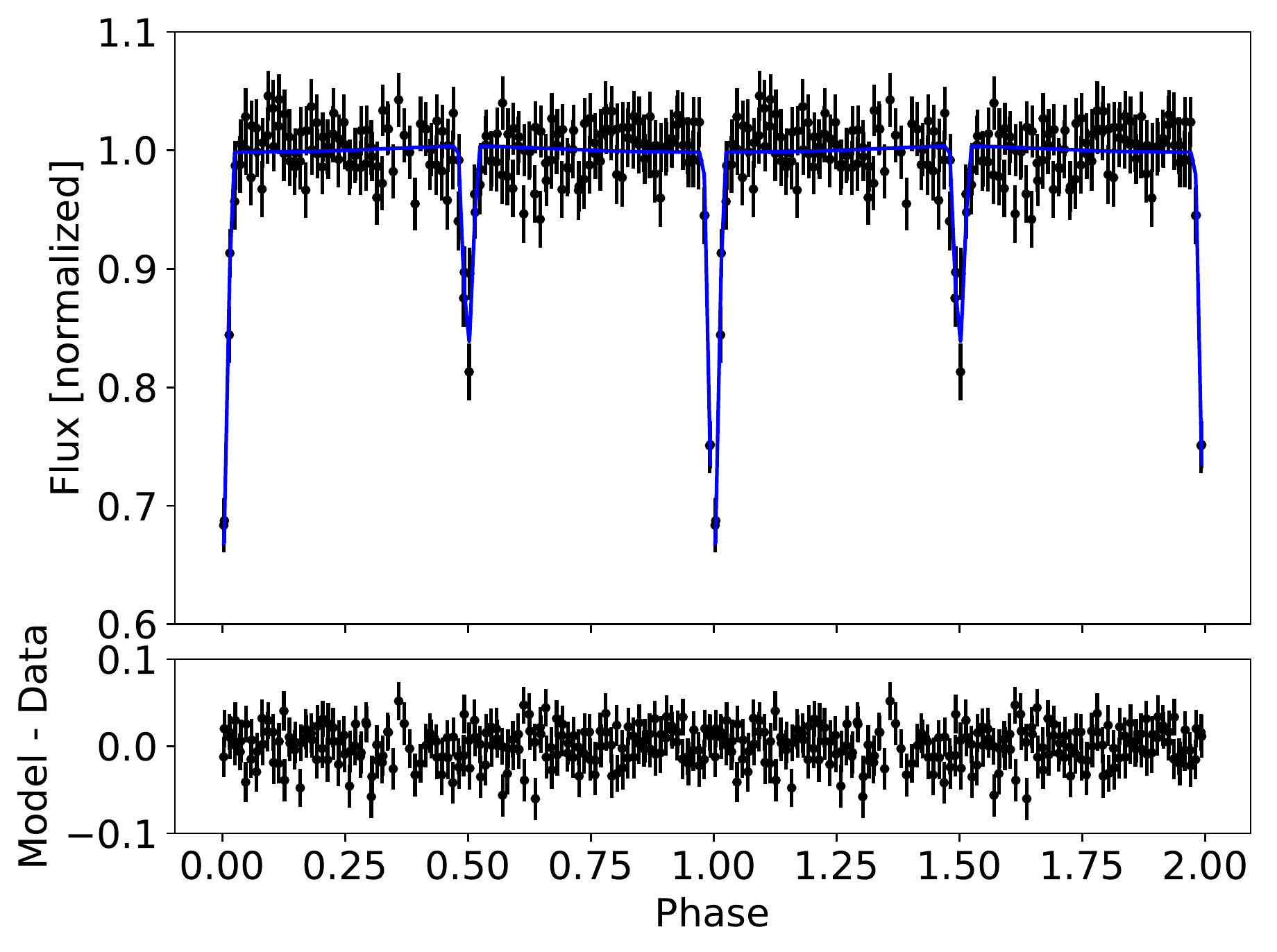}
 \caption{
On the left are the ZTF composite lightcurves of ZTF J190125.42+530929.5 (hereafter ZTF J1901+5309) with data taken between March and July 2018 (the approximate amount of data used for the identification step). On the right is the confirmation photometry from KPED observations in $g^\prime$, taken on the nights of 2019-03-04, 2019-03-18, 2019-10-03 and 2019-11-16; these sets of 1\,s duration observations lasted $\sim$\,2\,hr each. The blue trace is the best fit model.} 
 \label{fig:lightcurves}
\end{figure*}

The data from the current generation of astronomical surveys, with a variety of cadences, sensitivities, and fields-of-view, are resulting in an incredible variety of timescales to study variable stars.
These surveys include the Panoramic Survey Telescope and Rapid Response System (Pan-STARRS, \cite{MoKa2012}), the Optical Gravitational Lensing Experiment (OGLE, \cite{SoUd2013}), the All-Sky Automated Survey for Supernovae (ASAS-SN, \cite{JaKo2018}), the Asteroid Terrestrial$-$impact Last Alert System (ATLAS, \cite{ToDe2018}) and the Zwicky Transient Facility (ZTF, \cite{Bellm2018,Graham2018,MaLa2018}) among others.
The study of variable stars is an important part of understanding stellar and galactic structure and evolution. 
For example, some classes of variable stars such as Cepheids and RR Lyrae stars are standard candles and intrinsically bright, making them useful tracers of galactic structure \citep{SkSk2019}.
Due to the cadence at which many previous surveys have returned to the same location on the sky ($\geq 1$\,day), relatively little is known about the short-timescale variable optical sky. 
Below this nominal cadence, repeated observations of the same field at different phases of potential periodic sources allow for potential identification of variability at much shorter timescales.

Confirmation and characterization of the shortest-period variables require follow-up by dedicated resources.
For example, the Kitt Peak Electron Multiplying CCD (EMCCD) demonstrator (KPED), which is an EMCCD and filter wheel combination on the Kitt Peak 84 inch telescope, is designed for rapid and sensitive photometric follow-up of these sources \citep{CoDe2019}. A combination of its low-noise characteristics ($<$1\,$e^-$ rms), rapid readout speeds ($\approx$\,8\,Hz), and field of view ($4.4 \times 4.4$\,arcmin$^2$ field of view), is able to resolve low-amplitude variability on timescales of seconds. 

One example of potentially short-period, periodic sources are white dwarf (WD) stars, which arise from the final stage of stellar evolution for stars with initial masses below approximately 7-9\,$M_\odot$ \citep{DoNa2006}.
Future gravitational-wave detector missions in the millihertz band will detect WD binary systems; short-period and non-interacting binary systems make for excellent verification sources for these detectors \citep{KuKo2018} because it is possible to monitor the orbital decay in the optical for comparison to general relativity and the data from these gravitational-wave observatories. Some of the shortest period examples of this class are the 12.7-min system SDSS J065133.338$+$284423.37 (known as J0651) \citep{BrKi2011,HeKi2012} and the 6.9-min system ZTF J153932.16$+$502738.8 (known as ZTF J1539$+$5027) \citep{BuCo2019}. 
Many of these systems will merge within a Hubble time due to the loss of angular momentum through gravitational-wave radiation. 
Detached systems in particular are important, as interacting systems complicate direct tests of general relativity.

\begin{figure}
  \includegraphics[width=3.2in]{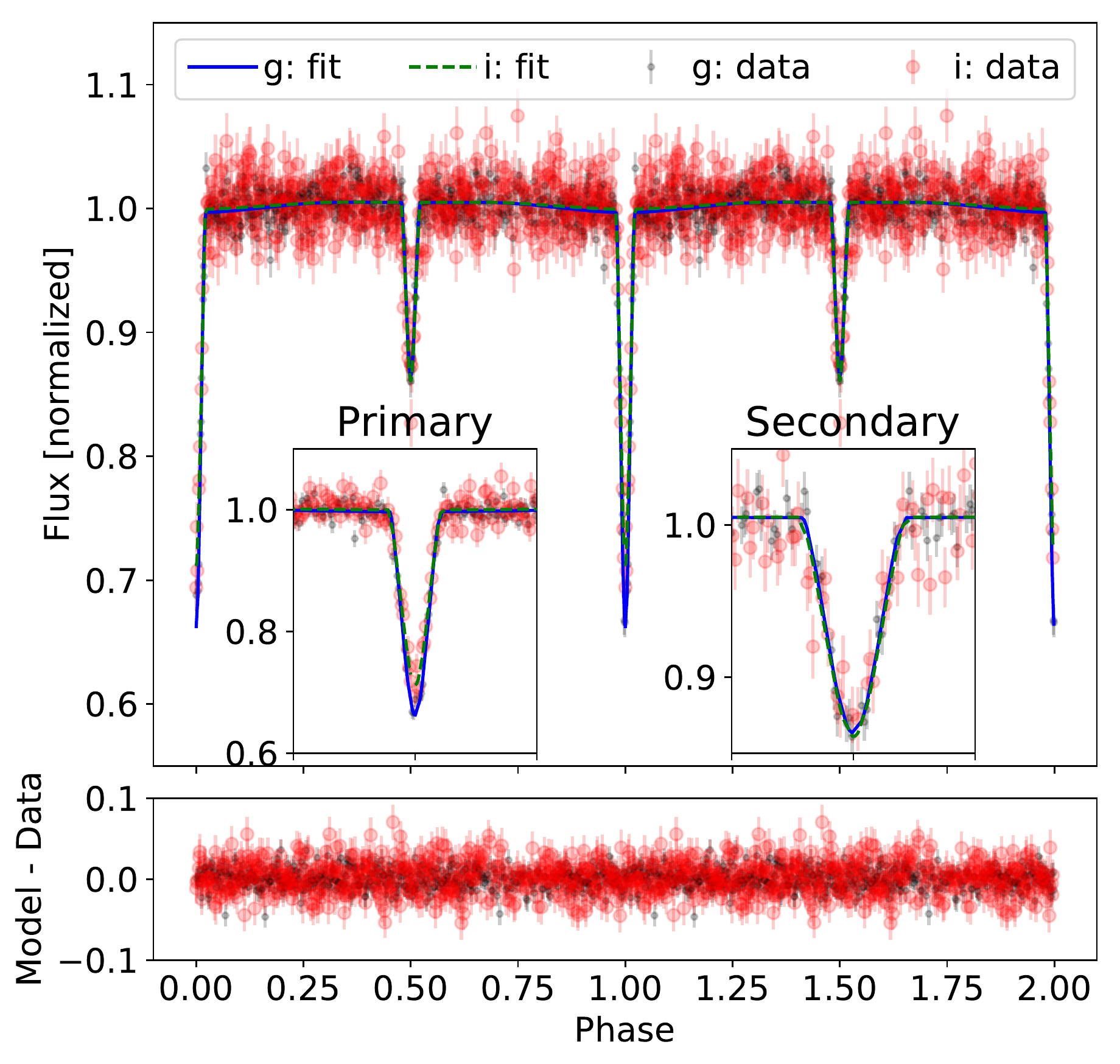}  
 \caption{
$g^\prime$ (black markers) and $i^\prime$ (red markers) lightcurves of ZTF J1901+5309 from CHIMERA \citep{HaHa2016}, the high-cadence photometer on the Palomar-200\,in, taken on the nights of 2019-04-29 and 2019-08-08. Here, we show 3\,s integrations beginning August 8, 2019 8:16:46 UTC and ending August 8, 2019 10:33:52 UTC. The blue and green traces are the best fit models for $g^\prime$ and $i^\prime$ respectively. We include insets showing zoomed versions of the primary and secondary eclipses as well.}
 \label{fig:lightcurves_CHIMERA}
\end{figure}

In this work, we highlight a 40.6-min period, white-dwarf binary system, ZTF J1901$+$5309, discovered by ZTF, with the eclipses and period confirmed by KPED.
The system most similar to the one presented here is SDSSJ010657.39$-$100003.3 (known as J0106$-$1000), which is a detached WD system with an orbital period of only 39.1 minutes, and was the first detection of a tidally distorted WD \citep{KiBr2011}.
ZTF J1901$+$5309 has magnitudes of $g=17.9$\,mag, $r=18.3$\,mag, $i=18.6$\,mag, and $NUV=17.3$\,AB\,mag (from Galex, \citealt{BiSh2017}), and from Gaia \citep{Gaia2018}, a parallax = $1.1 \pm 0.1$\,mas corresponding to $\rm D = 900^{+110}_{-90}\,pc$.
It has a right ascension of 19h 01m 25.42s and declination of 53$^\circ$ 09' 29.5'' (epoch J2000).
Section~\ref{sec:criteria} gives the selection criteria used to identify variable ZTF sources.
Section~\ref{sec:results} presents the observations for this system. 
We present our conclusions and paths for future work in Section~\ref{sec:conclusion}.

\section{Selection Criteria}
\label{sec:criteria}

%%Because our interest is in the population of white-dwarf binaries, we targeted blue objects, Pan-STARRS color $g-r < 0$. There are $\approx$\,10 million objects $\geq$ 50 available epochs (see \cite{BuCo2019} for further details).
%%When searching for periodic signals within the data set, periods were searched in the range from 3\,minutes up to $\sim$\,3 months.
%%In addition, we use a rolling-mean calculation to detrend the lightcurves, which combines to form a high-pass frequency filter and mean subtraction. This is optimized to find small amplitude, high-frequency variations.
%There are a variety of period finding algorithms in the literature \citep{ZeKu2009,MoFa2015,MoCo2017}. 

We discovered ZTF J1901+5309 by running the conditional entropy period-finding algorithm \citep{GrDr2013} over the ZTF observations of a sample of all Pan-STARRS objects satisfying the color cut $g-r<0.0$ and $r-i<0.0$, in the same manner as described in \cite{BuCo2019}. 
We selected a subset of these objects with more than 50 available ZTF epochs at the time, and combined the $g$, $r$, and $i$-band light curves by matching the median brightness values in each band. We required a minimum of 50 ZTF epochs in these light curves, which resulted in a sample of $\approx$\,10 million objects. 
We phase folded objects at their ``best'' period (that which corresponds to the minimum entropy), and saved thumbnails of each phase folded light curve. Then, by manual inspection, we selected systems to follow up (in particular, those which appeared to have a string of significant negative outliers phasing up into an eclipse-like feature). 
ZTF J1901+5309 was the second such object we identified after ZTF J1539+5027 \citep{BuCo2019}. 
This was possible due to its location in the ZTF high declination survey, which has a sampling cadence of 6 times per night (3 in ZTF $g$-band, 3 in ZTF $r$-band); this is the same survey in which ZTF J1539+5027 was discovered. 
As of the time of publication, ZTF J1901+5309 is consistently recovered as one of the highest significance periodic objects with a period under an hour using conditional entropy, and in searches using the box least squares algorithm \citep{Kovacs}, is the most significant periodic object we detect out of all white dwarf candidates identified in the Gentillo-Fusillo white dwarf catalog \citep{Fusillo2019}.

%%The algorithm used here, conditional entropy, is based on an information theoretic approach described in \cite{GrDr2013}.
%%We use the implementation in the \texttt{pycuda}-based \texttt{cuvarbase} package\footnote{ https://github.com/johnh2o2/cuvarbase} to perform the search.
%%The significance of a given lightcurve was determined from the algorithm; we keep significances above 7, corresponding to keeping 25\,\% of the lightcurves.

%For objects with high significances, the folded light curves are inspected for periodicity. Broadly speaking, folded lightcurves passing the significance criterion are either sinusoidal or sharply peaking.
%ZTF J1901$+$5309 has a significance of $\sim$\,29 in the search using the currently available lightcurves, \rednote{about 1\% of all objects}. We visualize more than 10,000 light curves and attention was drawn to this object because it showed evidence of being eclipsing.

\section{Observations}
\label{sec:observations}

On the left of Figure~\ref{fig:lightcurves}, we show the ZTF lightcurve available around the time of identification, phase-folding at the highest likelihood period (see Table~\ref{tab:Parameters}), plotting points at the mid-exposure time of ZTF's 30\,s exposures.
As of ZTF's second data release\footnote{https://www.ztf.caltech.edu/page/dr2}, there are 845 observations in ZTF $g$-band, 928 observations in ZTF $r$-band, and 62 observations in ZTF $i$-band.
The right of Figure~\ref{fig:lightcurves} shows the KPED lightcurve, confirming the object as an eclipsing binary. Circular, fixed-aperture photometry is performed on the KPED and CHIMERA images \citep{CoDe2019}. Differential photometry from a bright, nearby comparison star is used to remove transparency variations. 
Differential atmospheric extinction resulting in long-term photometric trends is removed using a low-order polynomial. All observations are converted from modified Julian date (MJD) in coordinated universal time (UTC) to barycentric Julian date (BJD) in barycentric dynamical time (TDB) using a light-travel time correction in $astropy$ \citep{PrSi2018} to account for the motion of the Earth around the barycenter of the solar system.

Our spectroscopic observations are from the DoubleSpec (DBSP) spectrograph of the Hale 200-inch telescope at Palomar Observatory on the night of 2019-05-29.
We obtained 8 epochs of spectroscopy of the system.
Using the 600/4000 blue grating with an angle at 27\,deg 17\,min and a 1.5'' slit resulted in 1\,\AA\,resolution and coverage from 3400-5700\,\AA.
At this resolution, exposures of 360\,s were possible, shown in Figure~\ref{fig:spectra}.
We also obtained seventy-two 180\,s exposures on the night of 2019-09-03 using the Low Resolution Imaging Spectrometer (LRIS) \citep{OkCo1995} on the 10-m W. M. Keck I Telescope on Mauna Kea, using the 400/8500 grism and $2 \times 2$ binning on the blue arm. 
The SNR in each observation is $\sim$\,10, with wavelength coverage from 3200-9500 $\si{\angstrom}$, and a resolution of approximately $\Delta \lambda / \lambda = 3000$.

\section{Results}
\label{sec:results}

\begin{table}
\centering
\caption{Table of physical parameters. $a$ is the orbital separation, $J$ is the surface brightness ratio of secondary to the primary, and $g$, the surface gravity, is in cgs units. The astrometry (epoch J2000), parallax, and distance measurements are taken from Gaia \citep{Gaia2018,BaRy2018}.} \label{tab:Parameters}
\begin{tabular}{c|c}
\hline
\hline
$\rm R_A$ & $0.07^{+0.01}_{-0.01}\,a$   \\
\hline
$\rm R_B$ & $0.08^{+0.01}_{-0.02}\,a$   \\
\hline
$\rm J$ & $0.42^{+0.02}_{-0.02}$   \\
\hline
$\rm i$ & $86.2^{+0.6}_{-0.2}\,\rm degrees $ \\
\hline
$T_{0}$ & $2458703.87381^{+0.00001}_{-0.00001}$\,$\rm BJD_{TDB}$ \\
\hline
q & $0.21^{+0.53}_{-0.15}$ \\
\hline
$\rm T_A$ & $28,000^{+500}_{-500}\,K$  \\
\hline
$\rm T_B$ & $17,600^{+400}_{-400}\,K$  \\
\hline
$\rm log(g)_{A}$ & $7.6^{+0.3}_{-0.3}$  \\
\hline
$P(T_{0})$ & $2436.11^{+0.02}_{-0.03}\,\rm s$  \\
\hline
\hline
$\rm RA$ & $285.35592397 \pm 0.00000003$\,deg   \\
\hline
$\rm Dec$ & $53.15815056 \pm 0.00000003$\,deg   \\
\hline
$\rm Parallax$ & $1.1 \pm 0.1$\,mas   \\
\hline
$\rm D$ & $915^{+96}_{-80}\,pc$   \\
\hline
\hline
\end{tabular}
\end{table}

A period search of the ZTF lightcurve shows an orbital period of P=40.6\,min. 
We do not see significant ellipsoidal variations, which indicates that the brighter object is not being significantly distorted due to rotation and tides, and thus is well within its Roche lobe.
We use the $g$-band CHIMERA lightcurve in Figure~\ref{fig:lightcurves_CHIMERA} to perform the lightcurve modeling.
Following the method from \cite{BuCo2019}, we use a \texttt{python} implementation of \texttt{Multinest} \citep{FeHo2009} to measure the radial-velocity covariances.
The sample lightcurves were generated using the \texttt{ellc} package \citep{Max2016}, where the lightcurves depend on the mid-eclipse time of the primary eclipse, $t_0$, the inclination, $i$, the mass ratio, $q=\frac{m_2}{m_1}$, the ratio of the radii to the semi-major axis, $r_{1}=R_{1}/a$, $r_{2}=R_{2}/a$, and the surface brightness ratio, $J$.
The first object is treated as spherical, the second is assumed to have a Roche geometry.
We use the period derived from the ZTF observations.
The resulting parameters from this fit are given in Table~\ref{tab:Parameters}.
We note that the mass ratio in the light curve modeling is very poorly constrained due to the lack of ellipsoidal modulation present.
There are some signs of one object being irradiated by the other; this effect can be useful in constraining the surface brightness ratio in such systems.
However, in an edge-on system with two eclipses such as this one, this quantity is much more strongly constrained by the relative depths of the eclipses.

\begin{figure}
\includegraphics[width=3.5in]{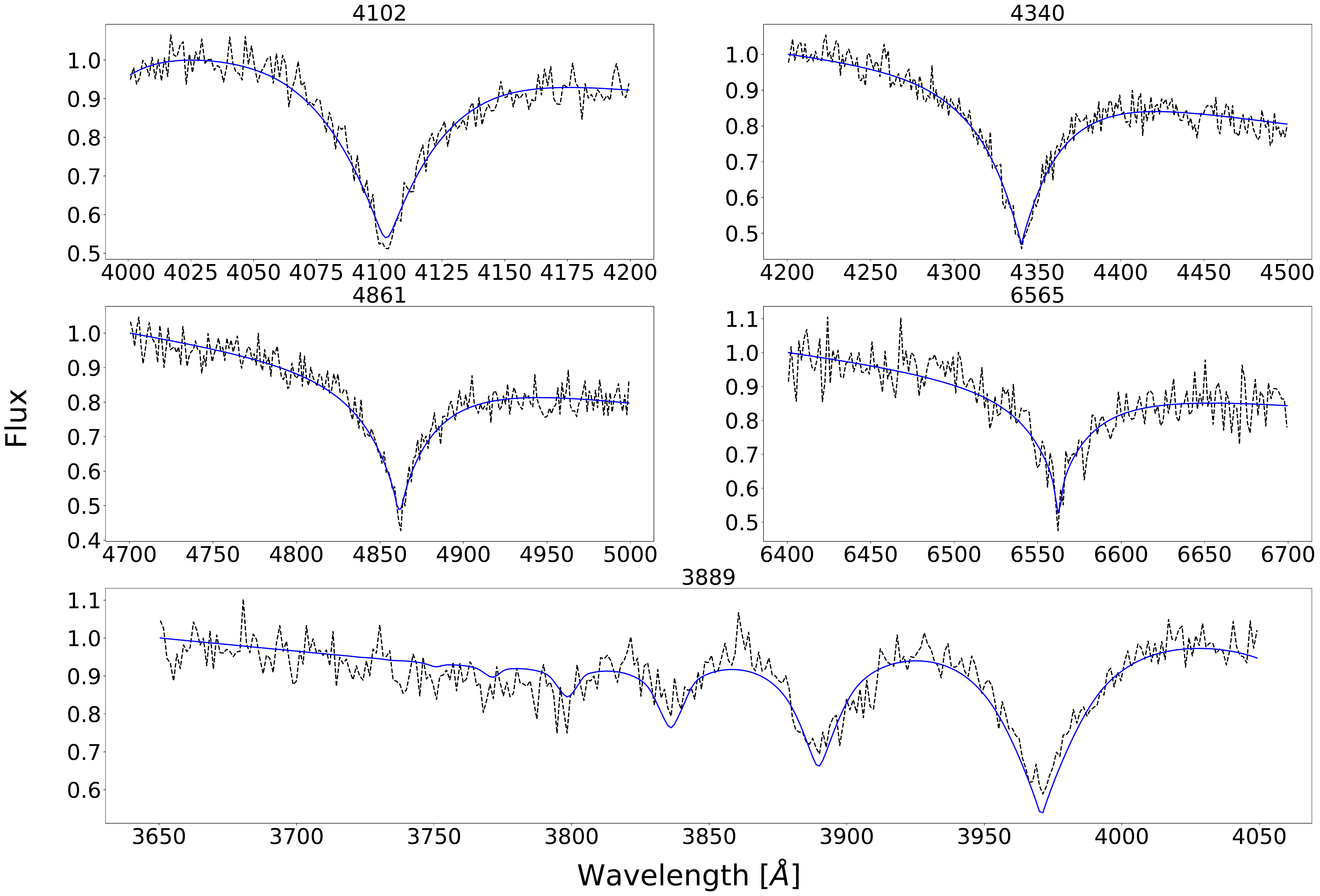}
\caption{Fits of the surrogate model based on the simulations of \citealt{LeDi2017} fit to the LRIS spectroscopy at orbital phase $\sim$\,0. This minimizes the artificial broadening of the lines as the white-dwarf spectra are superimposed, rather than separated. We highlight a handful of lines (the central wavelength shown in the plot titles) showing the spectral fits.}
\label{fig:fits}
\end{figure}

\begin{figure}
 \includegraphics[width=3.3in]{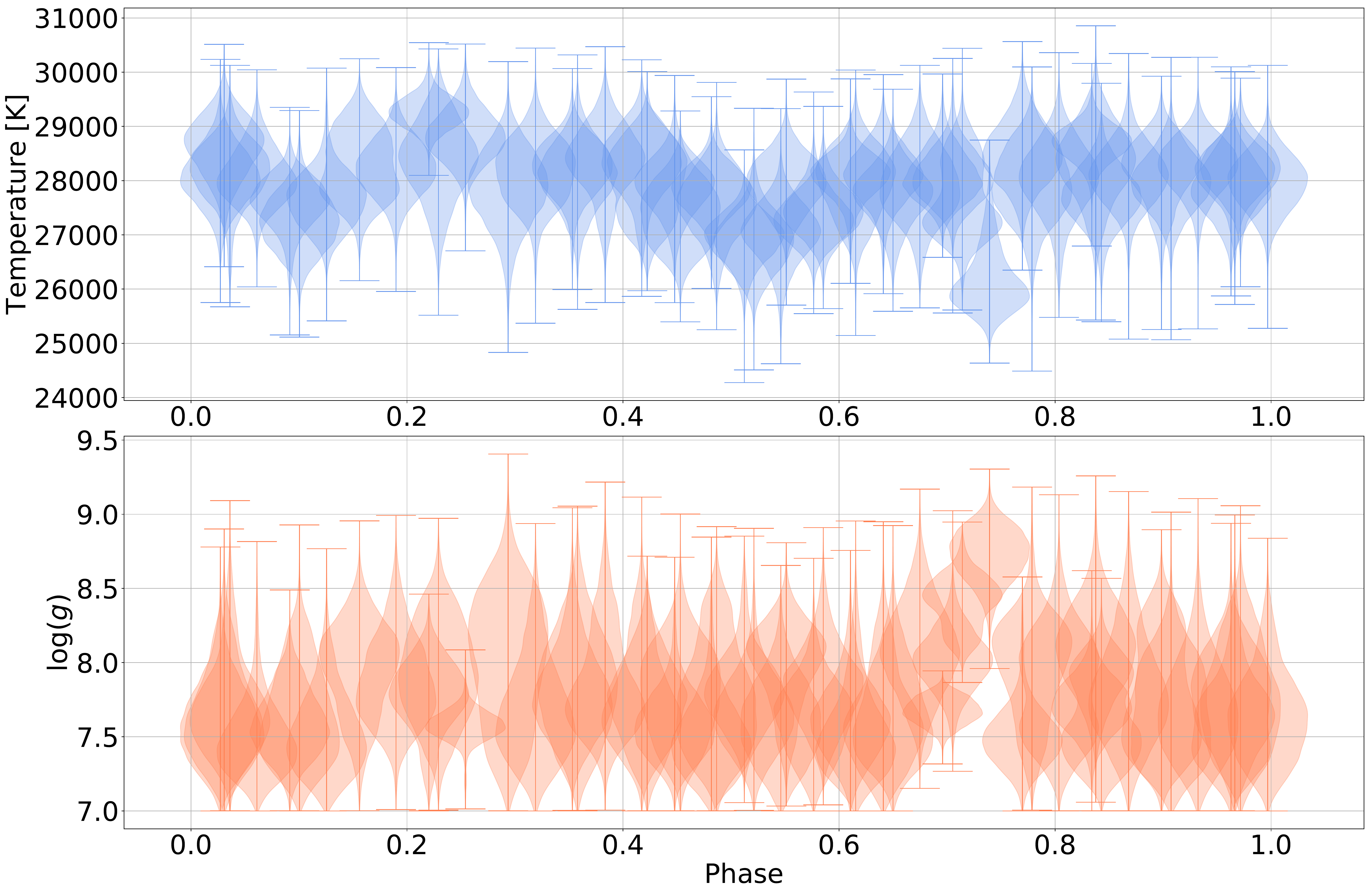}
 \caption{``Violin plot'' for the temperature and $\log g$ fits for LRIS spectra of ZTF J1901+5309 throughout its orbit.}
 \label{fig:T_logg}
\end{figure}

Since the spectra (Figure~\ref{fig:fits}) show broad Balmer lines of hydrogen from both white dwarfs, we use a library of synthetic spectra for white dwarfs with hydrogen-rich (DA) atmospheres from \cite{LeDi2017} to estimate the temperature and surface gravity.
The spectral grid covers the range $17,000 \textrm{K} \leq \textrm{T}_\textrm{eff} \leq 100,000 \textrm{K}$ and the surface gravity range $7.0 \leq \log(g) \leq 9.5$.
While the models only span a certain range of $\log(g)$ and helium abundance, as we observe DA white dwarfs, they should be reasonable to use.
We use a modification of a previously published, Gaussian Process Regression based technique~\citep{CoDi2018} to create surrogate spectral models to interpolate the spectra across the parameter space of interest.
We de-redden all of the spectra using the model of \cite{FiMa2007}.
Taking the spectrum nearest an orbital phase $\sim$\,0, we find the primary has $\textrm{T}_\textrm{eff} = 28,000 \pm 500$ and $\log(g) = 7.6 \pm 0.3$ in Figure~\ref{fig:fits}.
We find similar measurements across the rest of the DBSP as well as the LRIS spectra (see Figure~\ref{fig:T_logg}).
This implies a likely helium white dwarf, given its relatively low $\log(g)$ and large ratio of radius to semi-major axis at this orbital period, although it could also be consistent with a low mass carbon-oxygen core white dwarf.
Assuming both objects are approximately consistent with blackbodies in the observed passbands, and using the surface brightness ratio derived from the light curve modeling, we find the secondary has a temperate of $\rm T = 17,600^{+400}_{-400}\,K$.

We expect that these measurements may be biased by the potential contamination from the companion; we note that because the eclipse is not total, it includes a contribution from the secondary, and the overlap between the lines from the primary and secondary make joint fits difficult.
In particular, it is likely that $\log(g)$ may be overestimated.
We note in particular that the cooler companion is expected to have much wider ($\sim 2\,\times$) line widths than the primary, e.g. Fig. 10 of \cite{TeMa1966}, and so contamination of the wings by the companion is potentially significant. In addition, Stark broadening could be significant. Given that the pressure at a given optical depth varies roughly as $g^{2/3}$, e.g. Fig. 1 of \cite{TeMa1966} and Stark broadening has absorption coefficient $\alpha\propto
(p/T)\Delta\lambda^{-5/2}$, therefore, at a given optical depth in the wing, $\Delta\lambda\propto g^{4/15}$, $g \propto \Delta\lambda^{3.75}$. In this way, the $\sim$\,20\% variations in the line widths here could therefore give a spurious increase in the apparent $g$ by a factor of 2, or 0.3 in $\log (g)$.
As a point of comparision, we can use the Gaia \citep{Gaia2018} distance to make a black body radius estimate, with $F_\nu=R_A^2 \pi B_\nu(T)/D^2$. Using a black body temperature of 28,000\,K and the extinction-corrected $g$-band flux, we find $R_A = 1.85\times 10^9\mbox{cm}$; it is likely that such large radii are only attainable for hot white dwarfs of less than about $0.35M_\odot$ (or $\log g\lesssim 7.2$), see e.g. Fig 3. of \cite{PaAl2000}. 

Given that it is likely that the stars have already interacted in two common envelope phases, with potential mass exchange, evolutionary models do not apply here; however, single star evolutionary models, e.g. \citealt{CaAl2019,TrBr2011}, could in principle be used to guide plausible ranges of radii for both He and C/O WDs.

\begin{figure}
 \includegraphics[width=3.5in]{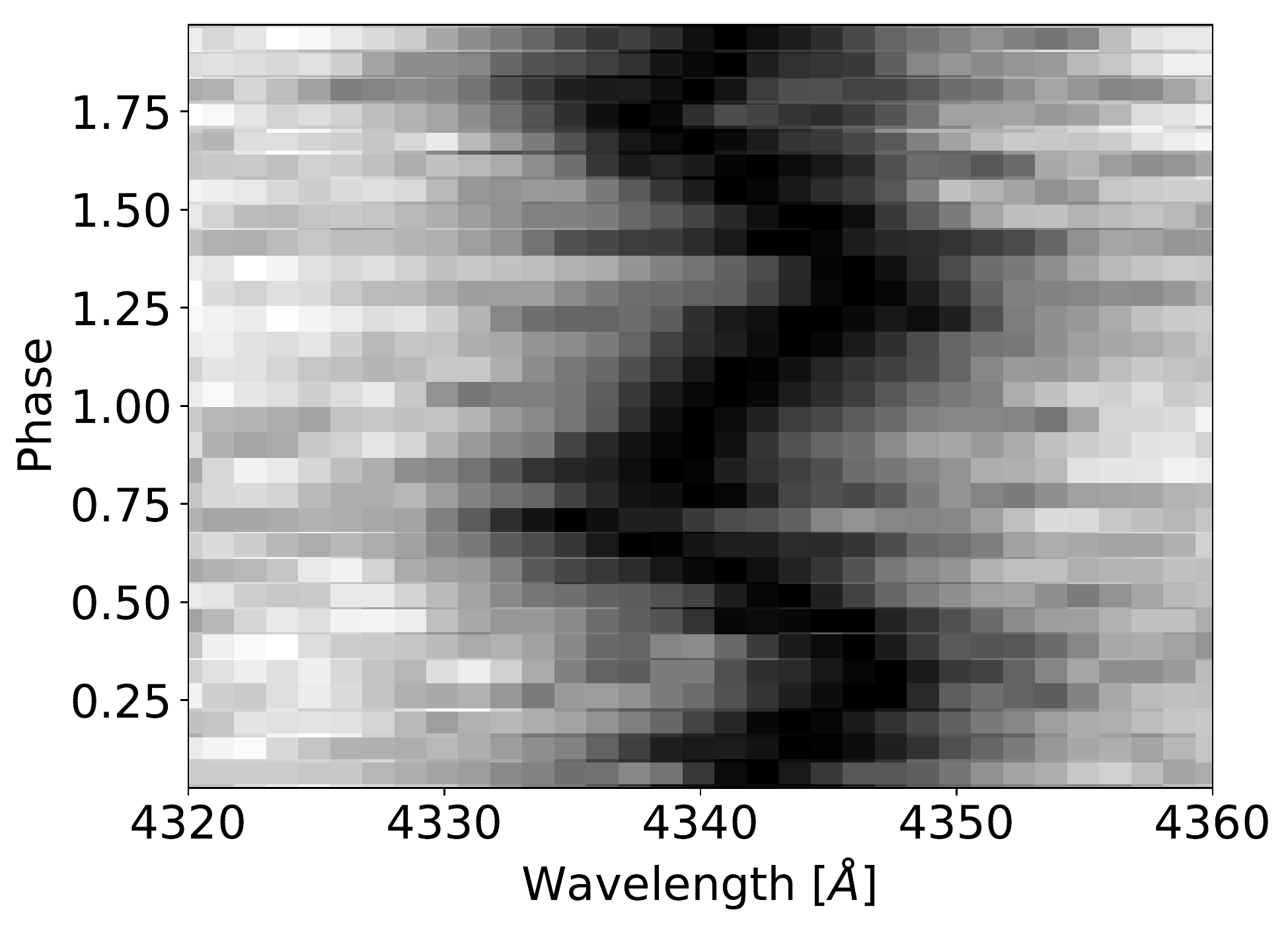}
 \caption{Array of LRIS spectra of ZTF J1901+5309 throughout its orbit focused on the H$\gamma$ transition at $\approx4340\,\si{\angstrom}$.}
 \label{fig:spectra}
\end{figure}

We now use the phase-resolved spectroscopy from LRIS to measure the orbital velocities of the white dwarfs in the binary (see \cite{BuCo2019} for a full discussion). 
The spectra exhibit broad and shallow hydrogen absorption lines, which are characteristic of a hot DA white dwarf, in Figure \ref{fig:T_logg}.
There are two overlapping absorption features here which move out of phase with one another; thus, this is a ``double-lined'' spectroscopic binary.
We once again use \texttt{Multinest} \citep{FeHo2009} to measure the radial velocity covariances.
As a simple model, we use Gaussians to fit the line associated with the $n=5$ to $n=2$ transition of hydrogen in each spectrum.
This yields a radial-velocity semi-amplitude of the objects, $K_1$ and $K_2$, which we fit to a sine wave
\begin{equation}
\begin{split}
v_i(n) = K_{\rm i} \sin \left(\frac{2 \pi n}{N}  + \phi_i \right) + A; \\
\end{split}
\label{eq:AB}
\end{equation}
where $n$ encodes the index of the phase-bin, $\phi_1 = 0$, and $\phi_2 = \pi$, $A$ is the systemic velocity and $v_i(n)$ is the velocity associated with $\phi_i$. 
We used a 1-dimensional Kernel Density Estimator (KDE) for each $K_{\rm i,measured}$ to assign the probabilities during sampling.
Unfortunately, due to the significant blending of the lines, the semi-amplitudes are significantly underestimated at 200\,km/s. For this reason, while the changes in radial velocity are clear (see Figure~\ref{fig:spectra}), it is not currently possible to extract quantitative measurements; while this does give a lower limit for comparison with future analyses, it gives an unphysically small mass function.
%\rednote{By eye of the line core in Fig 4, ESP gets a semi amplitude of 10/2=5 Angstroms=380km/s. Taking this core to be of the hotter white dwarf, and using Kepler's laws $v_{orb,A}=700\mbox{km s}^{-1}q(1+q)^{-2/3}$,
%so we infer $q=M_B/M_A=0.7$, which seems pretty reasonable. What is ESP missing?} 

\begin{figure}
\begin{center}
\includegraphics[width=3.5in]{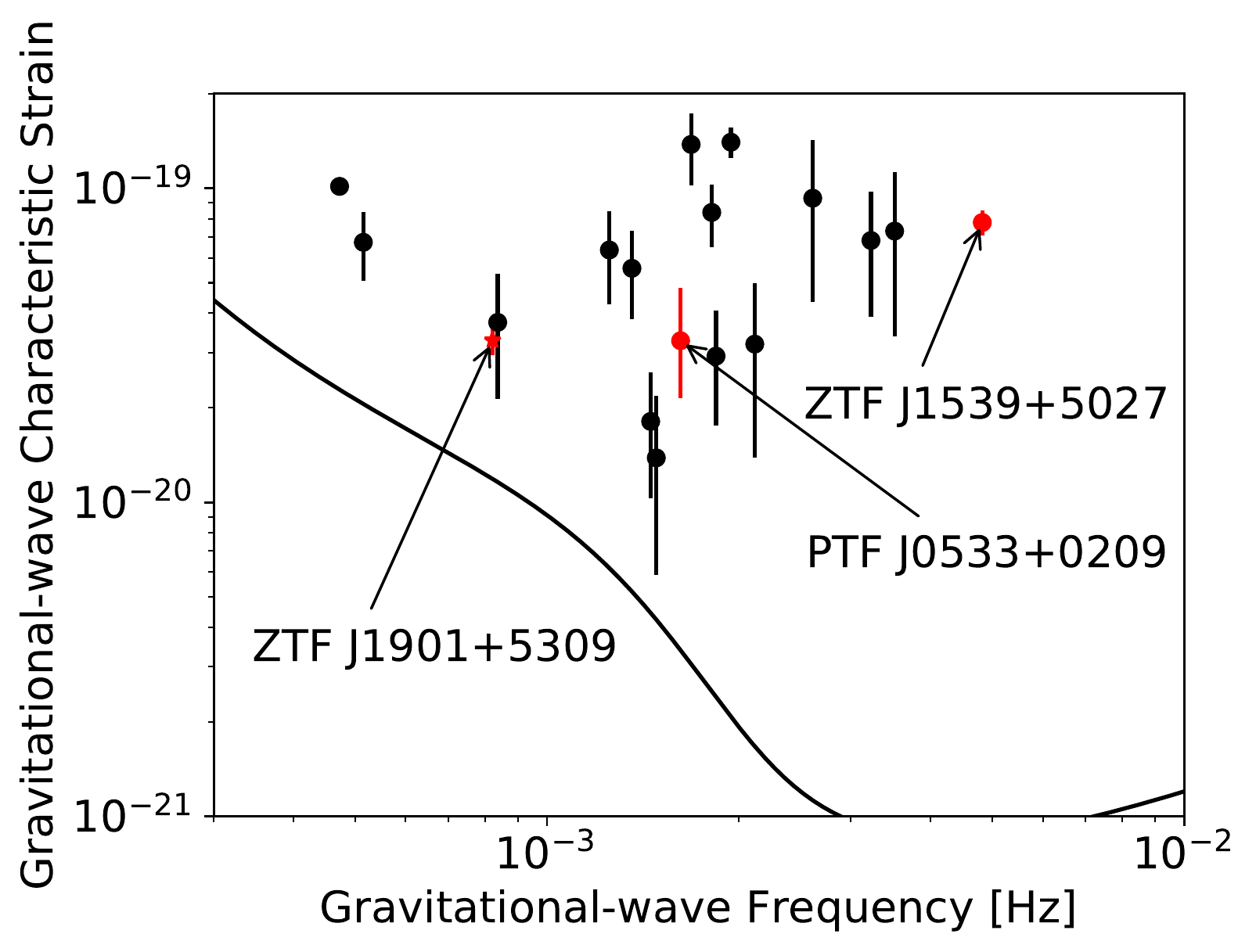}
\caption{Known \textit{LISA} detectable binaries assuming a 4-year observation time with the \textit{LISA} sensitivity curve in black. The previously known sources in black are taken from Kupfer et al. (2018), while the red star is ZTF J1901+5309 and the red circles indicate other sources from the PTF/ZTF search for \textit{LISA} sources (Burdge et al. 2019a,b).}
\label{fig:LISA}
\end{center}
\end{figure}

Given that we do not have a chirp mass, we take as a proxy a 0.5\,$+$0.3\,solar mass white dwarf system. With this assumption, knowledge of the orbital period, and the distance estimate from Gaia, we can estimate the gravitational-wave strain of the system. To calculate the characteristic strain \citep{KoRo2017}, $S_c$, we use
\begin{equation}
    \label{eq:strain}
   S_{c}=\frac{2(GM_{c})^{5/3}(\pi f)^{2/3}}{c^{4}D}\sqrt{fT_{obs}}
\end{equation}
where $c$ is the speed of light and $T_{obs}$ is the operation time of the \textit{LISA} mission. Marginalizing over the distance uncertainties, we provide the point estimate and uncertainties in Figure~\ref{fig:LISA}, along with other known \textit{LISA} sources. Accounting for the \textit{LISA}'s instrument response patterns \citep{KoRo2017}, the \textit{LISA} signal-to-noise ratio is around $2.4^{+0.3}_{-0.3}$ at the end of the nominal four-year mission lifetime, which would be on the edge of detectability.
For a 0.5\,$+$0.3\,solar mass white dwarf system, gravitational radiation will bring them into contact in approximately $2\times 10^7$\,years, when the orbital period will be around 5\,minutes.
The subsequent fate of the system after the onset of mass transfer depends sensitively on the (ill-determined) mass ratio, tidal coupling, and response to mass and angular momentum transfer \citep{MaNe2004,Shen2015}, and could end its life as a SN .Ia or a R Coronae Borealis star \citep{Shen2015}. 

\section{Conclusion}
\label{sec:conclusion}

In this work, we have described the discovery and characterization of the detached WD binary system ZTF J1901+5309, one of a growing number of systems that will be important sources for future space-based gravitational-wave detectors. We have illustrated the importance of high cadence photometric follow-up using target selection from ZTF to enable the robust identification and confirmation of short-period variables in particular. Using photometric and spectroscopic follow-up, we have constrained the WD mass ratio and scaled radii. Discovery of these systems is important for a variety of reasons. First of all, measurement of the space density and merger rate of WD binaries enable the test of whether they are the source of type Ia supernovae or other explosive events; e.g. Figure 3 of \cite{Shen2015}. 
In addition, there are a variety of predictions for the resulting evolution of sources of this type. For example, if the mass ratio is less than 0.6, stable mass transfer could lead to the creation of AM CVn systems like HM Cancri \citep{RoRa2010}. It is also possible that unstable mass transfer will cause the WDs to quickly coalesce and merge \citep{DaRo2011,Shen2015}. 

Going forward, follow-up high-cadence photometry will play an important role in confirmation of these systems and as a timing instrument.
By the time \textit{LISA} flies, high-cadence follow-up will allow for the measurement of the time derivative of the orbital period with time for many systems which show eclipses and have periods that are less than 1\,hour.
These objects require regular monitoring over the course of months to years to measure a change in period, depending on the masses and initial period (at discovery) of the objects.
There are ongoing observations to determine the orbital decay rate for this system. 

In addition, this system in particular will benefit from higher resolution spectra in order to disentangle the radial velocity measurements. While a measurement of the rate of change of orbital period is possible, more straightforwardly, this would enable measurement of the system's chirp mass; without either of these measurements, it may remain unknown until LISA flies.

\section*{Acknowledgments}
MC is supported by the David and Ellen Lee Prize Postdoctoral Fellowship at the California Institute of Technology.
SRK thanks the Heising-Simon's foundation for supporting his research with ZTF.
Part of this work was carried out at the Jet Propulsion Laboratory, under contract with NASA.
ESP's research was funded in part by the Gordon and Betty Moore Foundation through Grant GBMF5076.

Based on observations obtained with the Samuel Oschin Telescope 48-inch and the 60-inch Telescope at the Palomar Observatory as part of the Zwicky Transient Facility project. ZTF is supported by the National Science Foundation under Grant No. AST-1440341 and a collaboration including Caltech, IPAC, the Weizmann Institute for Science, the Oskar Klein Center at Stockholm University, the University of Maryland, the University of Washington (UW), Deutsches Elektronen-Synchrotron and Humboldt University, Los Alamos National Laboratories, the TANGO Consortium of Taiwan, the University of Wisconsin at Milwaukee, and Lawrence Berkeley National Laboratories. Operations are conducted by Caltech Optical Observatories, IPAC, and UW.
 
The KPED team thanks the National Science Foundation and the National Optical Astronomical Observatory for making the Kitt Peak 2.1-m telescope available. We thank the observatory staff at Kitt Peak for their efforts to assist Robo-AO KP operations. The KPED team thanks the National Science Foundation, the National Optical Astronomical Observatory, the Caltech Space Innovation Council and the Murty family for support in the building and operation of KPED. In addition, they thank the CHIMERA project for use of the Electron Multiplying CCD (EMCCD).

\bibliographystyle{mnras}
\bibliography{references.bib} % if your bibtex file is called example.bib

% Don't change these lines
\bsp	% typesetting comment
\label{lastpage}

\end{document}